\documentclass[10pt, conference]{IEEEtran}

\pdfoutput=1

\usepackage{algorithm}  
\usepackage{algorithmic}   
\usepackage{amsmath}    
\usepackage{amssymb} 
\usepackage{array} 
\usepackage[tight,footnotesize]{subfigure}
\usepackage{color}
\usepackage{url}
\usepackage{float}         
\usepackage{graphics}                               
\usepackage{graphicx}   
\usepackage{multicol}         
\usepackage{multirow} 
\usepackage{xspace}

\newcommand{\basker}{\emph{Basker}\xspace}

\newcommand{\superlud}{{SuperLU-Dist}\xspace}
\newcommand{\superlum}{{SuperLU-MT}\xspace}
\newcommand{\pardiso}{{Pardiso}\xspace}
\newcommand{\klu}{{KLU}\xspace}

\newcommand{\lnd}{{nested-dissection ordering}\xspace}
\newcommand{\nd}{{ND}\xspace}
\newcommand{\lamd}{{approximate minimum degree ordering}\xspace}
\newcommand{\amd}{{AMD}\xspace}
\newcommand{\lmatching}{{maximum weight-cardinality matching ordering}\xspace}
\newcommand{\matching}{{MWCM}\xspace}
\newcommand{\lbtf}{{block triangular form}\xspace}
\newcommand{\btf}{{BTF}\xspace}

\newcommand{\lgp}{{Gilbert-Peierls algorithm}\xspace}

\newcommand\verysmallfont{\fontsize{8}{9}\selectfont}

\newcommand{\mc}{{manycore processors}\xspace}

\newcommand{\str}[1]{\ensuremath{\mathcal{#1}}\xspace}

\begin{document}

\title{Basker: A Threaded Sparse LU Factorization Utilizing  Hierarchical Parallelism and Data Layouts}

\author{\IEEEauthorblockN{Joshua Dennis Booth}
\IEEEauthorblockA{Sandia National Laboratories\\
Albuquerque, New Mexico\\
jdbooth@sandia.gov}
\and
\IEEEauthorblockN{Sivasankaran Rajamanickam}
\IEEEauthorblockA{Sandia National Laboratories\\
Albuquerque, New Mexico\\
srajama@sandia.gov}
\and
\IEEEauthorblockN{Heidi K. Thornquist}
\IEEEauthorblockA{Sandia National Laboratories \\
Albuquerque, New Mexico,\\
hkthorn@sandia.gov}
}

\maketitle

\begin{abstract}
Scalable sparse $LU$ factorization is critical for efficient numerical simulation
of circuits and electrical power grids.  
In this work, we present a new scalable sparse direct solver called \emph{Basker}.
\emph{Basker} introduces a new algorithm to parallelize the \lgp for sparse $LU$ factorization.
As architectures evolve, there exists a need for algorithms that are
hierarchical in nature to match the
hierarchy in thread teams, individual threads, and vector level parallelism.
\basker is designed to map well to this hierarchy in architectures.
There is also a need for data layouts to match multiple
levels of hierarchy in memory. \basker
uses a two-dimensional hierarchical structure of sparse matrices that maps to
the hierarchy in the memory architectures and to the hierarchy in parallelism.
We present performance evaluations of \basker on the Intel SandyBridge and Xeon Phi
platforms using circuit and power grid matrices taken from the University of Florida
sparse matrix collection and from Xyce circuit simulations.
\basker achieves a geometric mean speedup of
$5.91\times$ on CPU (16 cores) and $7.4\times$ on Xeon Phi (32 cores) relative to KLU.
\basker outperforms Intel MKL Pardiso (PMKL) by as much as $53\times$ on CPU
(16 cores) and $13.3\times$ on Xeon Phi (32 cores) for low fill-in circuit matrices.
Furthermore, \basker provides $5.4\times$ speedup on a challenging matrix sequence taken 
from an actual Xyce simulation.
\end{abstract}

\IEEEpeerreviewmaketitle

\section{Introduction}
\label{sec:intro}

Scalable sparse direct linear solvers play a pivotal role in the efficiency of simulation codes on many-core systems.
Current approaches process multiple columns with similar nonzero structure (supernodes) with threaded Basic Linear Algebra Subprograms (BLAS)~\cite{superlu}.
The approach of using BLAS with one-dimensional data layouts of these matrices may not be able to extract enough parallelism when the matrix has low fill-in or an irregular nonzero pattern, such as matrices generated by Simulation Program with Integrated Circuit Emphasis (SPICE).
Therefore, a new type of solver is needed that uses a hierarchical  structures to leverage fine-grain parallelism within the irregular nonzero pattern.
In this work, we present a new shared-memory sparse direct $LU$ solver, \basker, designed to use hierarchical data layouts that exposes fine-grain parallelism and naturally fit the hierarchical memory structure of most many-core systems.
\basker is targeted towards parallelizing the state-of-the-art \lgp~\cite{gilbertlu} for low fill-in problems and thereby becoming the first parallel shared-memory solver to do so.
 
Sparse factorization of unsymmetric indefinite systems is difficult due to the need for numerical pivoting for stability and dynamic nonzero structure generated by such pivoting.
Scaling sparse $LU$ therefore depends on efficiently finding concurrent work inside this dynamic nonzero structure while providing enough numerical stability.
As a result, speedups achievable for sparse factorization is far from ideal~\cite{mumps, pardiso1}.
Coefficient matrices with low fill-in are particularly difficult, since the existence of supernodes is limited.
However, a hierarchical  structure can often be found in these matrices that can expose multiple levels of parallelism.

\basker uses a hierarchy of two-dimensional sparse blocks designed to exploit the nonzero structure that can be found in a matrix from circuit/powergrid problems.
These blocks can be found using traditional ordering techniques, such as \lbtf~\cite{klu} and  \lnd~\cite{scotch}.
This hierarchy of two-dimensional sparse blocks design allows \basker to accomplish two goals:
(1) exploiting any fine-grained parallelism found within or between blocks and
(2) designing a hierarchical data structure that fits the multiple levels of memory hierarchy and divide data among threads appropriately.
As a result, \basker enables parallelization of \lgp by allowing multiple threads work simultaneously on a single matrix column.

In this work, we present the algorithm and data layouts used by \basker to achieve hierarchical parallelism.
\basker is a implemented in templated C++11 with Kokkos~\cite{kokkos}. Kokkos is a package providing portability across multiple \mc and device backends. 
The main contributions of this work are:
\begin{itemize}
\item Parallelization of the Gilbert-Peierls algorithm;
\item A method to expose hierarchical parallelism in sparse matrices using two dimensional data-layouts;
\item A new threaded sparse direct $LU$ solver that outperforms Intel MKL's \pardiso~\cite{pardiso1} and \klu~\cite{klu} while reducing memory usage on matrices with low fill-in;
\item Empirical evaluation of \basker, KLU, and Pardiso on the Intel Sandy Bridge and Xeon Phi architectures.
\item Performance evaluation with $1000$ matrices from a transient simulation performed by the Xyce circuit simulator
\end{itemize}

The remainder of this paper is organized as follows.
Section~\ref{sec:background} presents an overview of previous solver work.
We then introduce the hierarchically structured algorithm to extract parallelism from sparse matrices in Section~\ref{sec:alg}.
Implementation choices are outlined in Section~\ref{sec:impl}.
Section~\ref{sec:results} provides performance results and comparisons with
other solvers.  Finally, possible future improvements and a summary of our
findings are described in Section~\ref{sec:conclusion}.

\section{Background and Related Work}
\label{sec:background}

This section provides a brief overview of background and related work to the solution of the sparse linear system $Ax=b$, where $A$ is a large sparse coefficient matrix, $x$ is the solution vector, $b$ is the given right-hand side vector.

\noindent\textbf{Orderings.}
All sparse direct solvers use structural information to improve performance and scalability.
Coefficient matrices are often reordered to limit fill-in, i.e., zeros becoming nonzero during factorization, or cluster nonzeros into patterns that reveal dependencies in computation.
Minimum degree orderings, such as \lamd (AMD), are a type of ordering that is very efficient in reducing fill-in~\cite{amd}. 
Nested-dissection (ND)~\cite{scotch} is another ordering based on the graph($G$) corresponding to a matrix, using $G(A)$ when $A$ is symmetric and $G(A$+$A^{T})$ when $A$ is unsymmetric. It is commonly used to provide a tree-structure  that can be used in parallel factorizations while reducing fill-in.

If an unsymmetric matrix does not have the strong Hall property, i.e., if every set of $k$ columns has nonzeros in at least $k$+$1$ rows, then the matrix can be permuted into a \lbtf (\btf) where block submatrices in the lower triangular part are all zeros.
A coefficient matrix $A$ permuted by matrices $P$ and $Q$ into \btf has the form:

{\verysmallfont
\[
PAQ = 
\begin{bmatrix}
A_{11} & A_{12} & \cdots & A_{1k} \\
       & A_{22} &        & \vdots \\
			 &        & \ddots & \vdots \\
			 &        &        & A_{kk}
\end{bmatrix} .
\]
}

This form is common in irregular unsymmetric systems, such as those from circuit simulation~\cite{klu}.
In this form, only submatrices on the diagonal ($A_{ii}$) need to be factored resulting in far less work, reduced memory usage, and a great deal of parallelism.
In addition to fill reduction, permuting the matrix to limit pivoting by placing nonzeros on the diagonal is common before computation~\cite{superludist}.
Finding such a permutation is done through finding a maximum cardinality matching of a bipartite graph representation of the coefficient matrix~\cite{duffmatching}. 
However, nonzeros on the diagonal is only one half of the issue; a variant that also tries to maximize the values on the diagonal is often used.
We will call this variant \lmatching (\matching)~\cite{duffmatching}.

\noindent\textbf{Sparse LU.}
We consider three popular solver packages, namely \superlud~\cite{superludist}, \pardiso~\cite{pardiso1}, and \klu~\cite{klu}, to compare their design choices to \basker.

\superlud is a distributed memory unsymmetric direct solver ~\cite{superludist}
that uses a two-dimensional data layout and avoids pivoting by using \matching that maximizes the sum of the diagonal element (MC64)~\cite{duffmatching}.
In each block matrix, \superlud performs a supernodal based $LU$ factorization.
However, supernodal methods have limitations such as
a pivot can only be chosen from inside a single supernode, fill-in must be known before hand, and scaling is limited by the size of supernodes~\cite{superlumt}.
A shared-memory version \superlum~\cite{superlumt} that uses a one-dimensional data layout exists.

\pardiso~\cite{pardiso1} is a shared-memory, supernodal, sparse $LU$ solver that uses a number of techniques to achieve high performance.
These techniques include using a left-right looking strategy to reduce synchronization and provide three levels of parallelism, namely from the etree, hybrid (left-right) at top levels, and pipelining parallelism. 
We compare against Intel MKL version of \pardiso and \superlum in Section~\ref{sec:results}.

\klu~\cite{klu} is a serial direct solver, based on the \lgp, and the closest
to our effort in algorithmic terms.
It achieves good performance by permuting the matrices first into \btf.
It then uses the \lgp to discover the nonzero pattern due to fill-in during numeric factorization in time proportional to arithmetic operations (Algorithm~\ref{alg:gp}~\cite{gilbertlu}).
However, \klu has no method to factor any part in parallel.
\basker was designed to replace \klu for circuit simulation problems by adding parallel execution both between blocks and within blocks of the \btf.
It is part of Trilinos library through both Amesos2~\cite{amesos2} and ShyLU~\cite{shylu} packages.


\begin{algorithm}[tbh]
  {\verysmallfont
    \begin{algorithmic}[1]
      \STATE //Input $A$.  Output $L,U,P$.
      \FORALL{$k$ columns of A}
        \STATE Find topological order of fill-in pattern of A(:,k) with depth-first search $\rightarrow$ $pattern_{k}$
        \FORALL{$j \in pattern_{k}$}
          \STATE $A(j+1:n, k) = A(j+1:n,k) - L(j+1:n,j)A(j,k)$
        \ENDFOR
        \STATE Select pivot from $A(:,k) \rightarrow P(k)$
        \STATE Copy from $A(:,k)$ to $L$ and $U$ based on $P$
      \ENDFOR
    \end{algorithmic}
  }
  \caption{Gilbert-Peierls Algorithm}
  \label{alg:gp}
\end{algorithm}

The primary features of \basker are:
(1) It is a nonsupernodal factorization;
(2) It uses a hierarchical data layouts;
(3) It uses both \matching and pivoting;
(4) It is a templated C++ solver using a manycore portable package supporting multiple backends such as OpenMP and PThreads.

\section{Basker Algorithm}
\label{sec:alg}
\begin{figure}[htbp]
  \centering
  \subfigure[]
     {
       \includegraphics[width=0.22\textwidth]{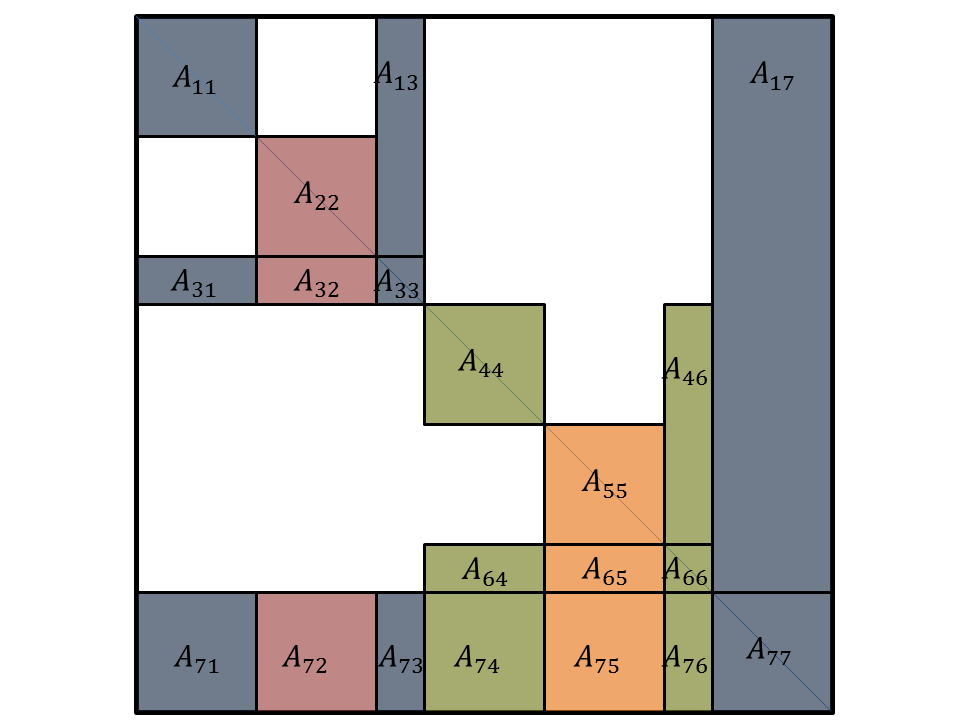}
       \label{fig:onedlayout}
     }
  \subfigure[]
  {
		\includegraphics[width=0.23\textwidth,height=0.16\textwidth]{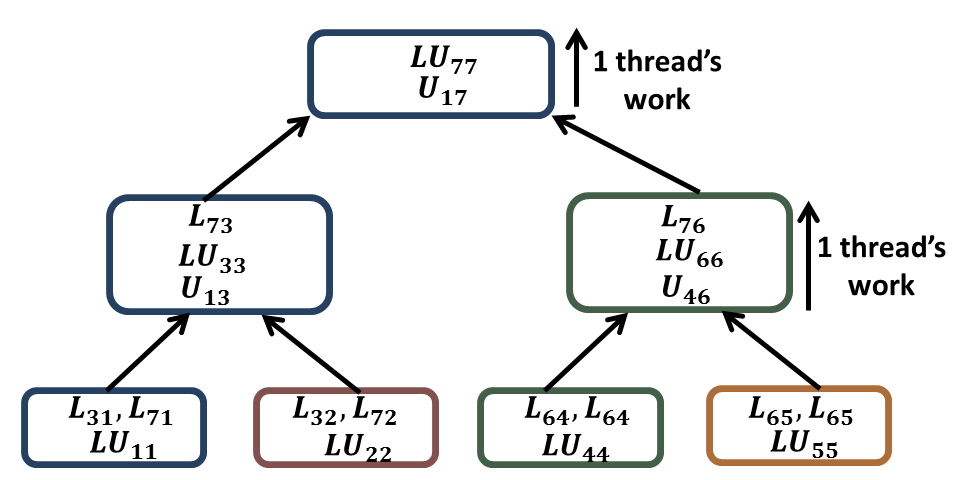}
    \label{fig:onedtree}
  }
  \caption{(a) One-dimensional layout of an ND-structure/binary $etree$ structure. The block $[A_{17} A_{77}]$ limits performance. The coloring provides one assignment of threads to computation. (b) Dependency tree of one-dimensional layout.  Note the large top level nodes that must be factored by one thread.}
  \label{fig:oned}
\end{figure}

This section introduces the parallel symbolic (pattern only) and numeric
(pattern and values) factorization algorithms in \basker.
The nonzero pattern of the coefficient matrix and the data-layout,i.e., how matrix 
entries are stored, determines not only the work but
also the available parallelism to a sparse factorization. 
Serial/multithreaded $LU$ 
factorization codes traditionally utilize a flat one-dimensional (1D) layout of blocks where blocks contain nonzeros in rows/columns stored contiguously.
These blocks are derived from some ordering of the matrix (e.g., See Figure~\ref{fig:onedlayout}).
However, using only 1D layouts limit the algorithms from exploiting sparsity
patterns within and between block structures.  
For instance, a 1D multithreaded supernodal factorization's speedup will be
 limited by the threaded BLAS on a set of columns (rows) called separators
(e.g, the block column $7$ in Figure~\ref{fig:onedlayout}). 
When these columns are not
dense, like in circuit/powergrid problems the use of BLAS is limited leading
to a serial bottleneck in the separators.
Due to this observation, \basker uses a variety of reordering
methods, such as \btf and ND, to derive a hierarchy of two-dimensional sparse
blocks.  This reordering allows \basker to fit the irregular nonzero pattern
into a hierarchy of blocks that fit the memory structure of modern nodes
and allow an algorithm that can utilize the 2D layouts (called 2D algorithm).
2D algorithms break columns into multiple submatrices 
(e.g., See Figures~\ref{fig:btf_both},\ref{fig:nd})
allowing
for multiple threads to work on a column that would have been serial in a
nonsupernodal method or efficiently use multiple calls of serial
BLAS.

In this work, we will focus on two levels of structures, i.e., 
\btf and ND.  We leave the third level (supernodes) within the 2D blocks
for future extensions.
\btf provides both the coarse structure for the whole matrix, \emph{and} the fine structure for a collection of submatrices.
\nd provides the fine structure for very large submatrices from BTF.
The fine structure of ND is used to arrive at a parallel 2D \lgp.

The notation used in this section is as follows.
A submatrix is given as $A_{ij}$, where $i$ and $j$ are the indices in the row and column in the two-dimensional block structure.
The nonzero pattern of a column ($c$) in a submatrix $A_{ij}$ is given as $\str{A}_{ij}(c)$. We use C++ notation for comments in the algorithms.

\subsection{Coarse Block Triangular Structure}
\begin{figure}[htbp]
	\centering
	\subfigure[]
	{
		\includegraphics[width=0.22\textwidth]{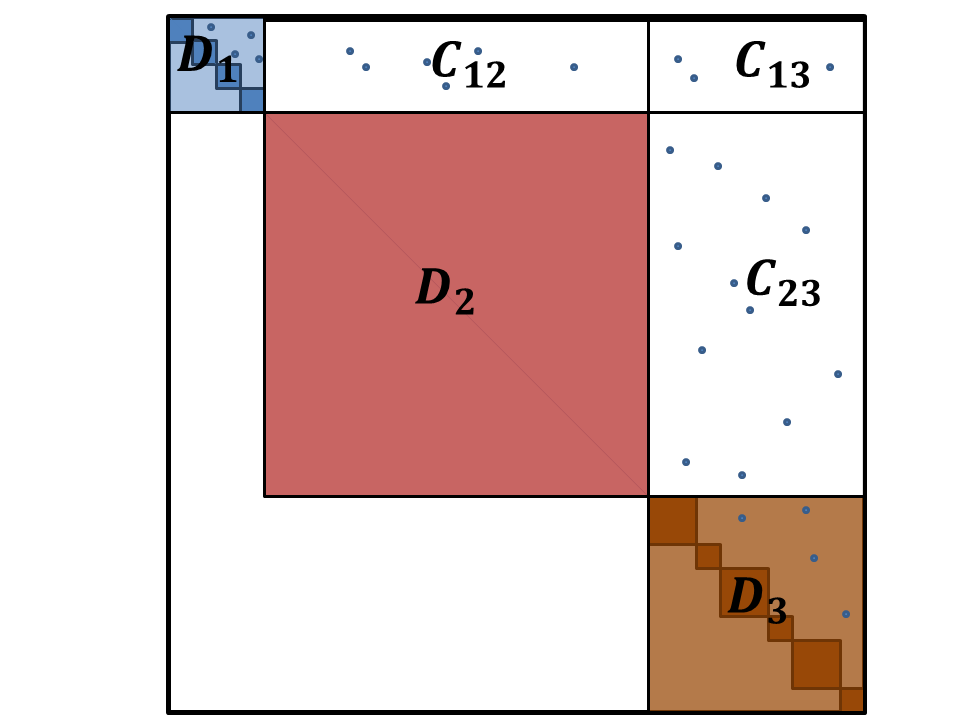}
		\label{fig:btf}
	}
	\subfigure[]
	{
		\includegraphics[width=0.22\textwidth]{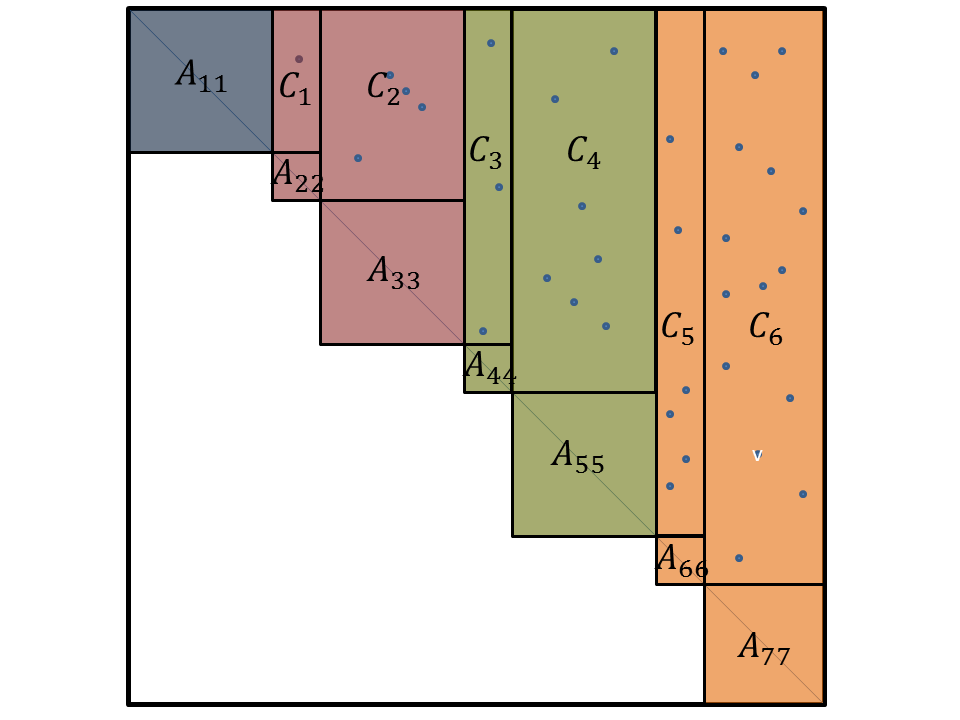}
		\label{fig:lower}
	}
	\caption{(a) Coarse structure, BTF ($P_{c}P_{m}AP_{c}^{T}$). The first level allows \basker to reduce factorization work by only factoring the diagonal blocks. (b) Representation of fine BTF structure, i.e., $D_{1}$ and $D_{3}$. Coloring of the blocks suggest one possible mapping of thread and blocks. } 
    \label{fig:btf_both}
\end{figure}

\basker uses \lbtf (BTF) on the input matrix to compute a coarse structure. 
It permutes the matrix based on an
ordering found from \matching ($P_{m1}$) to ensure a non-zero diagonal with
large entries.
A strongly connected components algorithm is used next to reorder the matrix ($P_{c}$) such that each component corresponds to a 
block diagonal. 
The reordered matrix, i.e., $P_{c}P_{m1}AP_{c}$, produces a structure similar to that in Figure~\ref{fig:btf}.
This form is common to matrices from several domains, and is well studied~\cite{pothenfan}.
Any of the large diagonal blocks may or may not exist for a particular matrix.

In Figure~\ref{fig:btf}, a two-dimensional structure with three diagonal blocks is shown.
As the multiple tiny subblocks in $D_{1}$ and $D_{3}$ provide enough natural
parallelism (for factoring each block), \basker uses this ordering derived from BTF as their second level structure as well.
The submatrices from this second level structure are handled using a
\emph{Fine Block Triangular Structure} based method.
In contrast, $D_{2}$ is very large without an opportunity to expose parallelism.
We will use ND to reorder $D_{2}$ further and use \emph{Fine Nested-Dissection Structure} based method.

\subsection{Fine Block Triangular Structure}
A typical representation of fine BTF structure, such as $D_{1}$ and $D_{3}$, is given in Figure~\ref{fig:lower}.
The substructure is easily dealt with as the subblocks are independent of each other.
Therefore, the sparsity pattern and factorization of each subblock ($A_{ii}$) can be computed concurrently.
A two-dimensional sparse block structure is used here.  
The off-diagonal blocks are ``partitioned'' in a manner  to help the sparse matrix-vector multiplication when solving for a given right-hand side vector.
They could further be split, however
they tend to be very sparse as they retain the original nonzero pattern.

\noindent\textbf{Parallel Symbolic Factorization.}
The symbolic factorization algorithm for the fine BTF block is shown in Algorithm~\ref{alg:sfactsmall}. It is embarrassingly parallel over the blocks.
We reorder each diagonal submatrix using \amd (Line 2) for fill-reduction.
Next, we find the number of nonzeros of each column and estimate the number of
floating-point operations required to factor (Line 3).
Using the number of floating-point operations, \basker assigns the submatrices among the threads and memory for $LU$ factors can be allocated.
The colors in Figure~\ref{fig:lower} provides one such assignment for four threads.

\begin{algorithm}[tbh]
 {\verysmallfont
   \begin{algorithmic}[1]
     \FORALL{subblocks on diagonal ($A_{ii}$) \textbf{IN PARALLEL}}
         \STATE \textbf{Compute} AMD order on $A_{ii}$ $\rightarrow$ $P_{amd}$
         \STATE \textbf{Compute} column count and number of operations of $P_{amd}A_{ii}P_{amd}^{T}$ 
     \ENDFOR
     \STATE \textbf{Partition} subblocks equally among $p$ threads based on number of operations
     \FORALL{ $p$ threads}
       \STATE \textbf{Initialize} $LU$ structure 
     \ENDFOR
   \end{algorithmic}
} 
\caption{Fine BTF Symbolic Factorization}
\label{alg:sfactsmall}
\end{algorithm}

\noindent\textbf{Parallel Numeric Factorization.}
After symbolic factorization, the numeric factorization uses the same thread mapping to submatrices to call sparse $LU$ factorization using \lgp.
The algorithm is not shown as it is a simple parallel-for loop over the diagonal submatrices.

\subsection{Fine Nested-Dissection Structure}
A subblock, such as $D_{2}$ in Figure~\ref{fig:btf} could be too large
to be factored in serial as in the above \btf fine structure method.
This block could easily dominate the factorization time, but there is no simple
way to factor this block with multiple threads with natural ordering.
This block constitutes an average of $68.4\%$ of the total matrix size in our problem test suite (see Section~\ref{sec:results}). As observed before,
using a 1D layout (Figure~\ref{fig:onedlayout}) does not provide enough parallelism. Instead we reorder
this block even further into finer 2D blocks.
Using this structure,we design the first parallel \lgp so multiple threads can work on a single column.

\begin{figure}[htbp]
	\centering
	\subfigure[]
	{
		\includegraphics[width=0.35\textwidth]{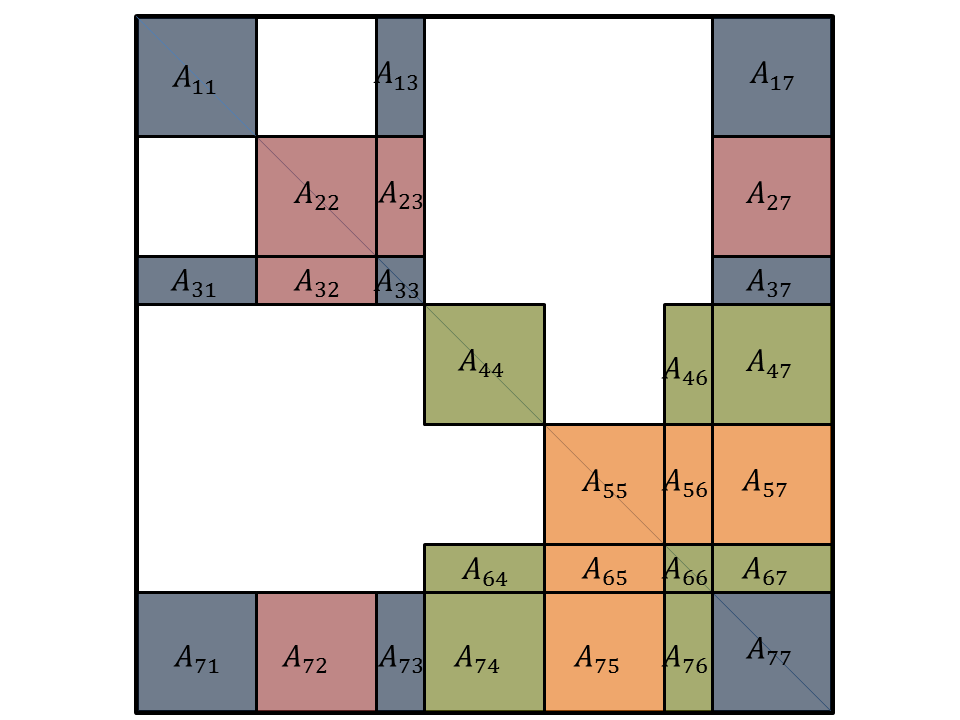}
		\label{fig:nd}
	}
\subfigure[]
{
		\includegraphics[width=0.35\textwidth]{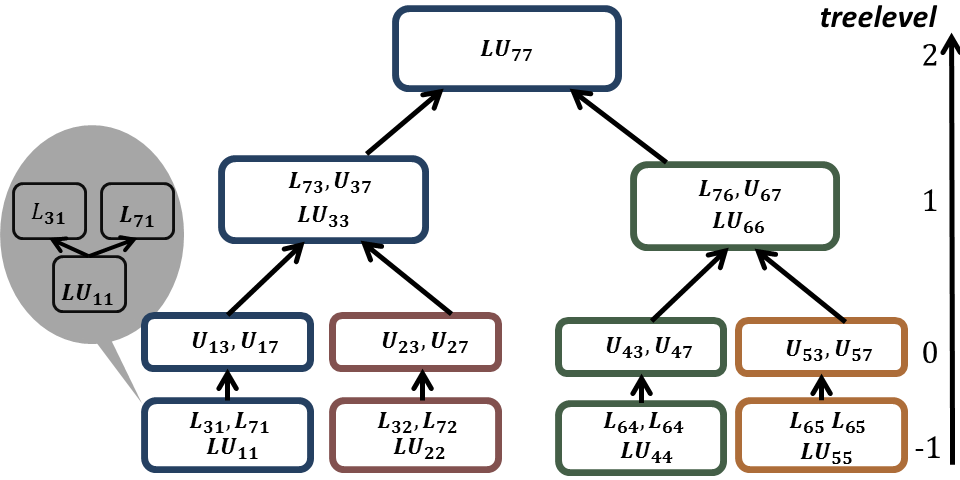}
		\label{fig:ndtree}
}
\caption{(a) Matrix in nested-dissection ordering of $D_{2}$. Each submatrix is stored by \basker as a sparse matrix. One possible thread layout indicated by color. Note, $LU$ will be stored in the same two-dimensional structure. (b) Dependency tree based off \nd structure.  The dependency is read from the bottom up, both within and between nodes.  The colors represent a static mapping of threads similar to those in (a).}   
\end{figure}

The \lnd is used in order to discover smaller independent subblocks to factor in parallel.
\basker first permutes $D_{2}$ using a \matching ($P_{m2}$) to find the locally best matching and reduce the need to pivot.
Next, \basker computes the \nd ordering on the graph of $D_{2}$+$D_{2}^{T}$ with a \nd tree.
\basker currently limits the number of leafs in the \nd tree to the number of threads available ($p$).
We note that increasing the number of leafs in the \nd tree may provide smaller cache friendly submatrices, but would limit the amount of pivoting allowed.
This trade-off is not explored in this paper. 
Additionally, current implementations of \nd provide only a binary tree, and therefore, \basker is limited to using a power of two threads.
The \nd ordering ($P_{nd}$) results in $P_{nd}P_{m2}D_{2}P_{nd}^{T}$, and the reordered matrix is given in Figure~\ref{fig:nd} for four threads.
This two-dimensional structure of sparse matrices is used to store both the reordered matrix and factorization ($LU$).
The colors suggest one possible layout where blocks of a particular color are shared by a thread.  

\noindent\textbf{Dependency Tree.} 
\basker requires a method to map the \nd structure to threads.
One option is to use a task-dependency graph, and use a tasking runtime.
However, \basker is currently limited to using data-parallel methods
(parallel-for) due to dependence on Kokkos and integration requirements with
Trilinos and Xyce.  \basker does this by
transforming a task-dependency graph into a dependency tree that represents
level sets that can be executed in parallel.

Figure~\ref{fig:ndtree} provides a general dependency tree used by both symbolic and numeric factorization for the two-dimensional matrix in Figure~\ref{fig:nd}, and is read from the bottom-up.
This tree represents two levels of dependency.
The first level dependencies are between matrices within a node. 
Within each node, matrices listed in a particular row depend on matrices listed in rows below in the same node.
For example, $L_{31}$ depends on having $LU_{11}$.
The second level dependencies are between nodes and are represented with arrows.
The levels in the dependency tree is denoted as $treelevel$, and $treelevel$ will always be used for only the dependency tree (not the $etree$ or \nd tree).
Nodes are colored to match the thread mapping in Figure~\ref{fig:nd}. 
Note that this tree is different from a \nd tree, and
expresses the concurrency in the hierarchical layout so \basker can use level scheduling.
One can easily see the difference with Figure \ref{fig:onedtree} where the root node
represents the entire $LU_{77}$ block column, whereas in the new dependence tree
$LU_{17} \ldots LU_{67}$ are distributed to multiple threads and the bottleneck
in the root node is much smaller.

\noindent\textbf{Parallel Symbolic Factorization.}
\basker now needs an accurate estimate of the nonzero count for the two-dimensional $LU$ factors found in parallel (Algorithm~\ref{alg:sfactlarge}).
A parallel symbolic factorization is crucial in a multithreaded environment as
repeated reallocation for $LU$ factors would require a system call,
which is a performance bottleneck when done in a parallel region. 
We do not form the $etree$ of the whole matrix and instead build
the appropriate portions of the $etree$ in different threads.

\basker first processes the bottom two levels in the dependency tree (Line 2-9) to obtain an accurate nonzero count.
The bottom most level of the dependency tree, i.e., $treelevel$ -1, has submatrices corresponding to $A_{11}, A_{22}, A_{33}, A_{44}$. First,
we find both the nonzero count per column and the $etree_{i}$~\cite{davisbook} of either $etree(A_{ii}$+$A_{ii}^{T})$ or $etree(A_{ii}A_{ii}^{T}$) (depending on symmetry and pivoting options) in parallel (Line 5).
Second, the nonzero counts for remaining $L_{ik}$ in the node at $treelevel$ -1 is found (Line 6). 
We note that $$\str{L}_{ik}(c) = \str{A}_{ik}(c)  \bigcup\limits_{t=1}^{c-1} \{  \str{L}_{ik}(t) | t \in \str{U}_{ii}(c) \}~\cite{rose}.$$
Also, pivoting while factorizing $A_{ii}$ will not affect $\str{ L}_{ik}(c)$ as $k > i$ by the fill-path theorem~\cite{rosetwo}. 
Therefore, \basker can use the above expression to find the nonzeros counts of the lower-diagonal submatrices.
Moreover, we find a data structure $lest$ with the maximum and minimum row index for each column $c$ that will be used for estimating nonzero counts in higher $treelevel$. 
At $treelevel$ 0, nonzero counts for the upper-diagonal submatrices, i.e., $U_{ki}$, can be found (Line 8). 
As $\str{U}_{ki}(c)$ may depend on the pivoting on $A_{ii}$ the $etree_{i}$ must be used. 
For each column ($c$), the method counts the nodes encountered starting from each nonzero in the column of $\str{A}_{ki}(c)$ to the least common ancestor of any nonzero already explored, where the least common ancestor of two nodes is the least numbered node that is the ancestor of both.
A data structure $uest$ is returned with the maximum and minimum row index for each row.


\begin{algorithm}[tbh]
{\verysmallfont
      \begin{algorithmic}[1]
        \STATE //treelevel = -1 to 0. Find $etree$ and nonzero count of submatrices on diagonal of lowest level in ND tree and lower half. 
        \FORALL{$p$ threads \textbf{IN PARALLEL}}
           \STATE \textbf{Map} $p \rightarrow i$
           \STATE //$treelevel$ = -1
	   \STATE \textbf{Compute} column count and $etree_{i}$ of $LU_{ii}$
           \STATE \textbf{Compute} column count of lower off-diagonal $L_{ki}$ $\forall k$ $\rightarrow lest_{k}$
           \STATE //$treelevel$ = 0
           \STATE Find column count of upper off-diagonal $U_{ik}$ $ \forall k$ $\rightarrow uest_{k}$
        \ENDFOR
	\STATE //Move up dependency tree
        \FORALL{$treelevel = 1:log_{2}(p)$}
            \FORALL{$nodes$ at $treelevel$ \textbf{IN PARALLEL}}
               \STATE \textbf{Map} $node \rightarrow j$
               \STATE \textbf{Compute} column count of diagonal submatrics corresponding to separators $LU_{jj}$ using $lest_{j}$ and $uest_{j}$
               \STATE \textbf{Compute} column count of lower off-diagonal submatrices corresponding to separators $L_{kj}$ using $lest_{k}$ and $uest_{j}$ $\rightarrow lest_{k}$ 
               \STATE \textbf{Compute} column count of upper off-diagonal submatrices corresponding to separator $U_{jk}$ using $lest_{j}$ and $uest_{k}$ $\rightarrow uest_{k}$
            \ENDFOR
        \ENDFOR
	\end{algorithmic}
	}
	\caption{Fine ND Symbolic Factorization}
	\label{alg:sfactlarge}
\end{algorithm}

The estimated nonzero counts for submatrices in the higher levels of the dependency tree are found using the estimates $lest$ and $uest$ by looping over the remaining $treelevels$ (Line 11).
At each $treelevel$, all the nodes on the level are handled by finding the nonzero count of the diagonal subblock, e.g., $LU_{33}$ (Line 14).
Now, 
$$\str{L}_{jj}(c) = \str{A}_{jj}(c) \bigcup\limits_{t=1}^{c-1} \{\str{L}_{jj}(t) | t \in \str{U}_{jj}(c) \} \bigcup\limits_{k=1}^{j} \str{L}_{jk}\str{U}_{kj}(c) $$
for these blocks, where $\str{L}_{jk}\str{U}_{kj}(c)$ is the pattern after the multiplication of $L_{jk}U_{kj}(c)$.
\basker estimates an upper bound of $\str{L}_{jk}\str{U}_{kj}(c)$ using the $lest$ and $uest$ by assuming the column is dense between the minimum and maximum if $lest$ and $uest$ overlap for the column.
We find that this is a reasonable upper bound and cheaper than storing the whole nonzero pattern.
Finally, the column count of any off-diagonal submatrices, such as $L_{73}$ and $U_{37}$, can be computed (Line 15 and 16). 
The column count for these submatrices use the upper bound as well (i.e., fill-in estimated with $lest$ and $uest$).

\begin{figure*}[!t]
	\centering
	\subfigure[]
	{
	
		\includegraphics[width=0.24\textwidth]{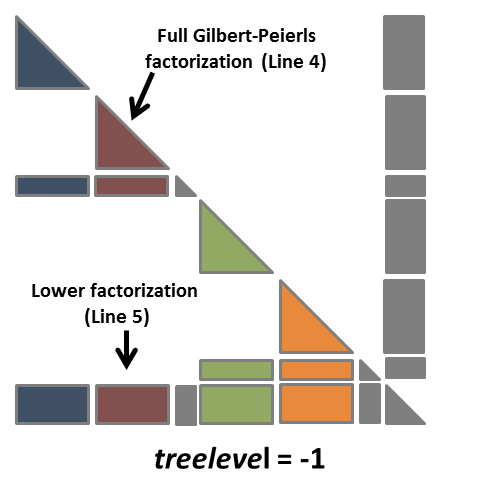}
		\label{fig:sone}
	}
	\subfigure[$slevel$ 1]
	{
		\includegraphics[width=0.08\textwidth]{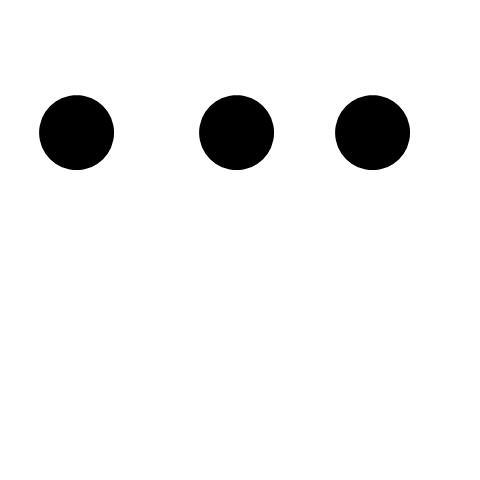}
		\label{fig:stwo}
	}
	\subfigure[Last separator column, $slevel$ 2]
	{
	
		\includegraphics[width=0.24\textwidth]{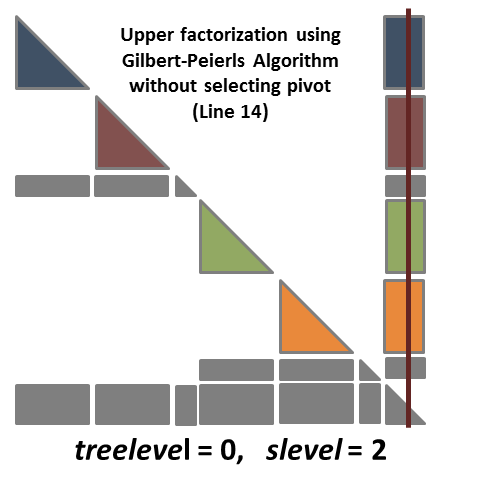}
		\label{fig:sthree}
	}
	\subfigure[]
	{
	
		\includegraphics[width=0.24\textwidth]{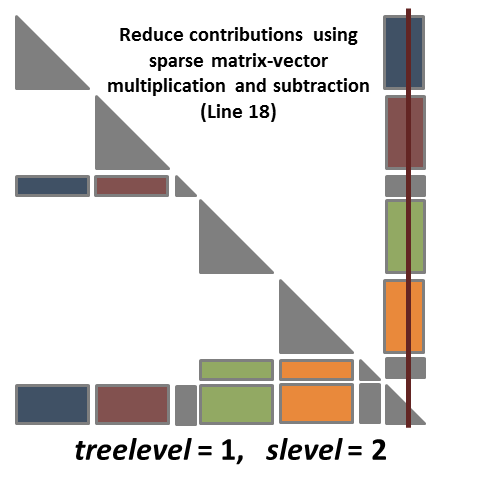}
		\label{fig:sfour}
	}
	\subfigure[]
	{
	
		\includegraphics[width=0.24\textwidth]{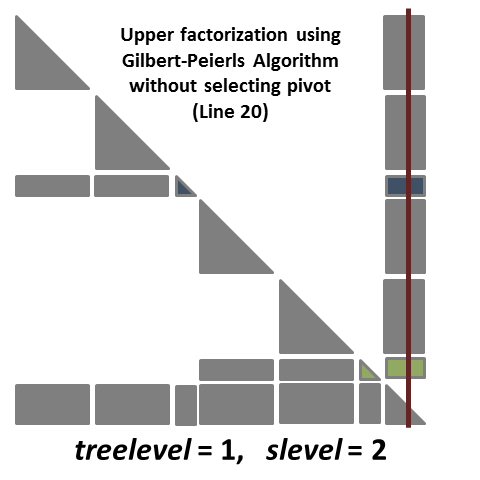}
		\label{fig:sfive}
	}
	\subfigure[]
	{
	
		\includegraphics[width=0.24\textwidth]{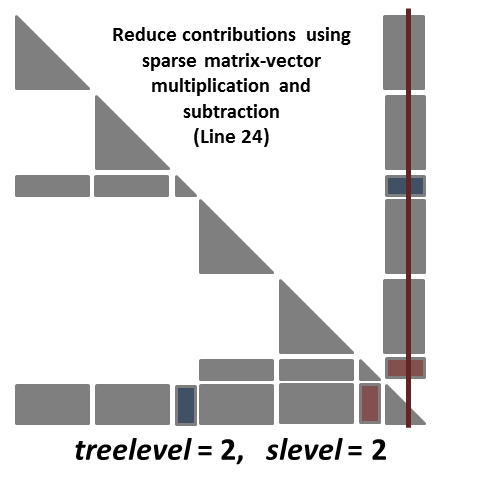}
		\label{fig:ssix}
	}
	\subfigure[]
	{
	
		\includegraphics[width=0.24\textwidth]{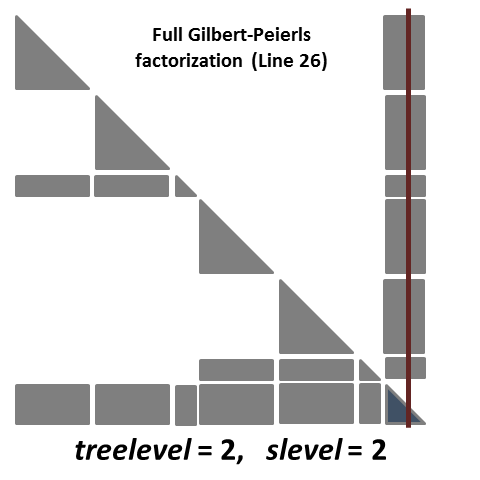}
		\label{fig:sseven}
	}
\caption{Workflow of Algorithm~\ref{alg:nfactlarge} (Numeric Factorization for Fine ND structure). The subblock forming the lower triangle are subblocks of $L$ and the red line indicates column of $A$ being factored. Note, the only serial bottleneck, i.e., a single colored block, is the bottom right most block in (g).}
		\label{fig:nfactorfig}
\end{figure*}

\noindent\textbf{Parallel Numeric Factorization.}
This subsection describes the parallel left-looking \lgp (Algorithm \ref{alg:nfactlarge}).
To facilitate understanding, we explain the algorithm using a series of block diagrams of the execution in Figure~\ref{fig:nfactorfig}.
Blocks that are not colored gray represent submatrices that are active/used at a stage, and the colors correspond to the thread mapping as in Figure~\ref{fig:ndtree}.  


\begin{algorithm}[tbh]
{\verysmallfont
  \begin{algorithmic}[1]
    \STATE //$treelevel$=-1 
    \FORALL{$p$ threads \textbf{IN PARALLEL}}
       \STATE \textbf{Map} $p \rightarrow i$ where $i$ is a leaf node
       \STATE \textbf{Factor} diagonal submatrices $A_{ii}$ $\rightarrow LU_{ii}$
       \STATE \textbf{Factor} lower off-diagonal submatrices $A_{ki}$ $\rightarrow L_{ki} \forall k$
    \ENDFOR
    \STATE \textbf{Sync} all threads
    \STATE //Factor remaining submatrices columns
    \FORALL{$slevel = 1:log_{2}(p)$}
       \STATE \textbf{Map} $slevel \rightarrow j$ 
       \FORALL{$p$ threads \textbf{IN PARALLEL}}
           \STATE \textbf{Map} $p \rightarrow i$ where $i$ is a leaf node
           \STATE //$treelevel$=0
           \STATE \textbf{Factor} upper off-diagonal submatrices $A_{ij}$ $\rightarrow U_{ij}$
	   \FORALL{$treelevel = 1:slevel$-$1$}
	      \STATE \textbf{Map} $sublevel \rightarrow l$
              \STATE \textbf{Sync} select threads
              \STATE \textbf{Reduce} contributions from previously found $U_{l_{1}j}$, $U_{l_{2}j}$ into upper off-diagonal submatrix $A_{lj}$ $\rightarrow \hat{A}_{lj}$
              \STATE \textbf{Sync} select threads
              \STATE \textbf{Factor} upper off-diagonal submatrices $\hat{A}_{lj}$ $\rightarrow U_{lj}$
	    \ENDFOR 
	    \STATE //$treelevel$=$slevel$, lower half of column
            \STATE \textbf{Sync} select threads
            \STATE \textbf{Reduce} contributions from previously found $U_{l_{1}j}$, $U_{l_{2}j}$ into diagonal submatrix $A_{jj}$ $\rightarrow \hat{A}_{jj}$
	    \STATE \textbf{Sync} select threads
            \STATE \textbf{Factor} $\hat{A}_{jj}$ $\rightarrow LU_{jj}$
            \STATE \textbf{Sync} select threads
	    \STATE \textbf{Factor} lower off-diagonal matrices $A_{kj}$ $\rightarrow L_{kj} \forall k$
	    \ENDFOR 
	    \STATE \textbf{Sync} all threads
	    \ENDFOR 
  \end{algorithmic}
}
\caption{Fine ND Numeric Factorization}
\label{alg:nfactlarge}
\end{algorithm}

Submatrices are factored based on the dependency tree in Figure~\ref{fig:ndtree} in a \emph{column-by-column} manner.
Figure~\ref{fig:sone} starts with the submatrices in $treelevel$ -1. 
\basker factors the submatrices on the diagonal that have no dependencies, i.e., computing $LU_{ii}(c)$ (Line 4).
This factorization uses the \lgp similar to Algorithm~\ref{alg:gp} in parallel on each
submatrix.
Next, the just computed column $U_{ii}(c)$ is used to compute column $c$ in the lower off-diagonal submatrices in the node at $treelevel$ -1, e.g., $L_{31}(c)$ and $L_{71}(c)$ (Line 5).
This is done by discovering the nonzero pattern as a result of parallel sparse matrix-vector multiplication.
At $treelevel$ -1, a level synchronization between all threads is needed 
before moving to next $treelevel$.
Note that \basker need not necessarily sync all threads if done in a task parallel manner. 

The nodes in the dependency tree starting at $treelevel = 0$ has a subtle
but important distinction.
All submatrices in a tree node are not computed before moving to next node
as in the symbolic factorization.
In contrast, only those submatrices in a tree node corresponding to a particular column $slevel$ are computed (Line 9). The $slevel$ indicates multiple
passes over the dependency tree (bottom up until $treelevel$).
Figures~\ref{fig:sthree}- \ref{fig:sseven} show the block diagram of $slevel=2$ with $treelevel=0,1, \mbox{ and } 2$, where the red line indicates the column being factored.
Submatrices at this $treelevel=0$ (Figure~\ref{fig:sthree}), e.g., $U_{17}$, are factorized in parallel using a method similar to Algorithm~\ref{alg:gp} except that $L_{ii}$ is used for the backsolve (Line 14).

\basker continues up the dependency tree with a loop over $treelevel$ (Line 15).
At each new level, \basker must synchronize specific threads in order to combine their results (Line 18).
Figure~\ref{fig:sfour} shows the blocks used in the reduction.
The reduction has two phases. 
The first phase is multiple parallel sparse matrix-vector multiplication of the matrices colored in $L$ and the column of $U(c)$ just found (the red line in the colored blocks).
The second phase is subtracting each threads' matrix-vector product from the corresponding blocks in $A$ (where gray blocks in the column are $A_{37}(c)$, $A_{67}(c)$ and $A_{77}(c)$).
For example, one thread computes the reduction results in $\hat{A}_{37}(c) = A_{37}(c) - L_{31}U_{17}(c) - L_{32}U_{27}(c)$. $\hat{A}_{67}(c)$ and $\hat{A}_{77}(c)$  are computed in
parallel as well.
Once the reduction is complete, the newly updated submatrix at $treelevel$ can be factored similar to other upper off-diagonal matrices (Line 20).
Figure~\ref{fig:sfive} provides a visual representation of this step.
At the last step, when $treelevel = slevel = 2$, at the root, there is one
reduction needed to the already computed $\hat{A}_{77}(c)$ (Line 24, 
Figure~\ref{fig:ssix}) and then a simple factorization in the diagonal block
can be computed
(Line 26, 
Figure~\ref{fig:sseven}). This last factorization is the \emph{only} serial
bottleneck in the algorithm.

In the more general case, when $treelevel$ = $slevel$ (Line 22) and we are not
at the root node (not shown in the figures), 
there is no farther bottom-up traversal of the dependency tree. This would have been true for the $treelevel = slevel = 1$ for block column three in our example.
In matrix terms, this means that $U(c)$ for a column has been computed
and only the block diagonal and $L$ remain to be computed (e.g, $L33(c)$,
$U_{33}(c)$, and $L_{73}(c)$). This requires a reduction (Line 24)
and factoring the diagonal submatrix (Line 26) as before, but
any lower off-diagonal submatrices of $L$ that remain, such as $L_{73}(c)$,
 need to be factored as well (Line 28).

\section{Basker Implementation}
\label{sec:impl}
\noindent\textbf{Data Layout.}
\basker uses a hierarchy of two-dimensional sparse matrix blocks  to store both the original matrix and $LU$ factors.
The 2D structure is composed of multiple compressed sparse column (CSC) format matrices.  
Parallelism must be extracted from between blocks in the \btf structure and within large blocks in order to achieve speedup on low fill-in matrices.
In particular, a hierarchical structure needs to be exploited to reveal more parallelism.
Additionally, this also breaks the problem into fine-grain data structures that better fit the structure of memory in modern many-core nodes.
\basker implements this by building this structure of C++ classes during the symbolic factorization after applying the aforementioned orderings.

\noindent\textbf{Synchronization.}
Light weight synchronizations are needed to allow multiple threads to work on a single column in \basker.
There are multiple places where these synchronizations need to happen in \basker, and they are marked in Algorithm~\ref{alg:nfactlarge}.
The number of threads that need to synchronize depends on location and iteration in the algorithm.
For instance, all threads need to synchronize moving from factoring leaf nodes and parent nodes, but only two threads need to sync in separator columns.
 
A traditional data-parallel approach launches parallel-for over a set of threads, and these threads rejoin the master only after the end of the loop.
However, if synchronization takes place between all threads at every level, the overhead would be too high.
In particular, the total time spent for synchronization in this manner for matrix G2 Circuit with 8 cores is 11\% of total time. 
Therefore, \basker uses a different mechanism to synchronize between threads.
This mechanism is a point-to-point synchronization that utilizes writing to a volatile variable where synchronization only happens between two threads that have a dependency.
Point-to-point synchronization's importance in the speedup of sparse triangular solve has been shown before~\cite{pointcom}.
Using this method, \basker is able to reduce synchronization overhead to 2.3\% ($\sim 79\%$ improvement) of total runtime for G2 Circuit.

\section{Empirical Evaluation}
\label{sec:results}
We evaluate \basker against Pardiso MKL 11.2.2 (PMKL), SuperLU-MT 3.0 (SLU-MT), and KLU 1.3.2 on a set of sparse matrices from circuit and powergrid simulations in terms of memory and runtime. 
Our MWCM implementation is similar to MC64 bottle-neck ordering~\cite{duffmatching}, unlike SuperLU-Dist's product/sum based MC64 ordering. 
Scotch~\cite{scotch} 6.0 is used to obtain the \nd ordering. 
Furthermore, we compare \basker's performance on a sequence of $1000$ matrices from circuit simulation of interest.

\subsection{Experimental Setup}


\begin{table}[tbh]
  \centering
	{\verysmallfont
	\tabcolsep=0.08cm
  \caption{Matrix Test Suite. $n$ represents dimension of matrix, $|.|$ is the number of nonzeros in the matrix. The minimum number of nonzeros between the factors of Basker and PMKL is in bold. * indicates Sandia/Xyce matrices, + indicates powergrids.}
    \begin{tabular}{c|c|c|c|c|c|c|c|c}
               &       &       &   KLU 		 & Pardiso   & Basker     & BTF      & BTF & KLU \\
    Matrix     &  $n$  & $|A|$ & $|L$+$U|$ & $|L$+$U|$ & $|L$+$U|$ & \% & blocks & $\frac{|L+U|}{|A|}$  \\
		\hline
    RS\_b39c30+ & 6.0E4 & 1.1E6 & 6.9E5 & 6.3E6 & \textbf{6.9E5} & 100 & 3E3   & 0.6 \\
    RS\_b678c2+ & 3.6E4 & 8.8E6 & 5.8E6 & 5.9E7 & \textbf{5.8E6} & 100 & 271   & 0.7 \\
    Power0*+       & 9.8E4 & 4.8E5 & 6.4E5 & 9.1E5 & \textbf{6.4E5} & 100 & 7.7E3 & 1.3 \\
    Circuit5M    & 5.6E6 & 6.0E7 & 6.8E7 & 3.1E8 & \textbf{7.4E7} & 0   & 1     & 1.3 \\
    memplus      & 1.2E4 & 9.9E4 & 1.4E5 & \textbf{1.3E5} & 1.4E5 & 0.1 & 23    & 1.4 \\
    rajat21      & 4.1E5 & 1.9E6 & 2.8E6 & 4.9E6 & \textbf{2.8E6} & 2   & 5.9E3 & 1.5 \\
    trans5       & 1.2E5 & 7.5E5 & 1.2E6 & 1.3E6 & \textbf{1.2E6} & 0   & 1     & 1.6 \\
    circuit\_4   & 8.0E4 & 3.1E5 & 5.0E5 & 5.8E5 & \textbf{5.1E5} & 34.8& 2.8E4 & 1.6 \\
    Xyce0*       & 6.8E5 & 3.9E6 & 4.7E6 & 3.8E7 & \textbf{4.8E6} & 85  & 5.8E5 & 1.8 \\
    Xyce4*       & 6.2E6 & 7.3E7 & 4.5E7 & 5.0E7 & \textbf{4.5E7} & 12  & 7.5E5 & 2.0 \\
    Xyce1*       & 4.3E5 & 2.4E6 & 5.1E6 & 5.6E6 & \textbf{5.1E6} & 21  & 9.9E4 & 2.4 \\
    asic\_680ks  & 6.8E5 & 1.7E6 & 4.5E6 & 2.9E7 & \textbf{4.5E6} & 86  & 5.8E5 & 2.6 \\
    bcircuit     & 6.9E4 & 3.8E5 & 1.1E6 & 1.1E6 & \textbf{1.1E6} & 0   & 1     & 2.8 \\
    scircuit     & 1.7E5 & 9.6E5 & 2.7E6 & 2.7E6 & \textbf{2.7E6} & 0.3 & 48    & 2.8 \\
    hvdc2+       & 1.9E5 & 1.3E6 & 3.8E6 & \textbf{3.0E6} & 3.8E6 & 100 & 67    & 2.8 \\
		\hline
		\hline
    Freescale1   & 3.4E6 & 1.7E7 & 7.1E7 & \textbf{5.6E7} & 6.8E7 & 0   & 1     & 4.1 \\
    hcircuit     & 1.1E5 & 5.1E5 & 7.3E5 & \textbf{6.7E5} & 7.1E5 & 13  & 1.4E3 & 6.9 \\
    Xyce3*       & 1.9E6 & 9.5E6 & 7.6E7 & \textbf{4.3E7} & 7.7E7 & 20  & 4.0E5 & 9.2 \\
    memchip      & 2.7E6 & 1.3E7 & 1.3E8 & \textbf{6.5E7} & 9.4E7 & 0   & 1     & 9.9 \\
    G2\_Circuit  & 1.5E5 & 7.3E5 & 2.0E7 & \textbf{1.3E7} & 2.0E7 & 0   & 1     & 27.7 \\
    twotone      & 1.2E5 & 1.2E6 & 4.8E7 & \textbf{2.7E7} & 4.7E7 & 0   & 5     & 39.9 \\
    onetone1     & 3.6E4 & 3.4E5 & 1.4E7 & \textbf{4.3E6} & 1.2E7 & 1.1 & 203   & 40.8 \\
		\hline
    \end{tabular}%
		}
  \label{tab:testsuite}%
\vspace{-0.3cm}
\end{table}%

\noindent\textbf{System Setup.}
We use two test beds for our experiments.  The first system has two eight-core
Xeon E5-2670 running at 2.6GHz (SandyBridge).  The two processors are
interconnected using Intel's QuickPath Interconnect (QPI), and share 24GB of
DRAM.  The second system has an Intel Xeon Phi coprocessor with 61 cores running
at 1.238GHz and 16GB of memory.
Since \basker requires a power of two threads, we only test up to 32 cores as 64 threads would oversubscribe the device. 
All codes are compiled using Intel 15.2 with
-03 optimization.

\noindent\textbf{Test Suite.}
\basker is evaluated over a test suite of circuit and powergrid matrices taken from Xyce and the University of Florida Sparse Matrix Collection~\cite{mat}.
These matrices vary in size, sparsity pattern, and number of BTF blocks.
Additionally, these matrices vary in fill-in density, i.e., $\frac{|L+U|}{|A|}$ where $|A|$ is the number of nonzeros in $A$.
We note that fill-in can be $< 1$ when using \btf, since only the diagonal subblocks of $A$ are factored to $LU$.
In Davis and Natarajan~\cite{klu}, coefficient matrices coming from circuit simulation generally have lower fill-in density than those coming from PDE simulations, i.e., $\frac{|L+U|}{|A|} < 4.0$.
Matrices with lower fill-in tend to perform better using a \lgp than a supernodal approach.
For fairness, we include seven matrices with fill-in density larger than $4.0$.
Table I lists all matrices sorted by increasing fill-in density measured using KLU.
The percent of matrix rows in small independent diagonal submatrices (Fine BTF Structure) is shown as BTF\%. The total number of BTF blocks is also shown.
A double line divides matrices with fill-in density higher than $4.0$. 
The test suite is
a mix of matrices with very different properties to exercise different portions of \basker.

\subsection{Memory Usage}
We now compare memory requirements in terms of $|L$+$U|$.
Table I lists the number of nonzeros in $L$+$U$ for KLU, PMKL, and \basker.
We do not report results for SLU-MT due performance considerations (shown below).
The nonzeros reported for PMKL and \basker are from a run using 8 cores on SandyBridge. We note that this number varies slightly for \basker depending on number of cores.
The best result between PMKL and \basker is in bold. 
We observe that \basker provides factors with less nonzero entries for most matrices with fill-in density $<$ 4.
This reduction can be as high as an \emph{order of magnitude} for the matrix RS\_b678c2+. This is the result of using the BTF structure and using fill reducing ordering on the subblocks.
However, PMKL uses slightly less memory on matrix with fill-in density $>$ 4.
The additional memory used by \basker on these matrices is far less than the additional memory used by PMKL on the first group of matrices.

\subsection{Performance}
\begin{figure}[htbp]
  \centering
  \includegraphics[width=0.49\textwidth]{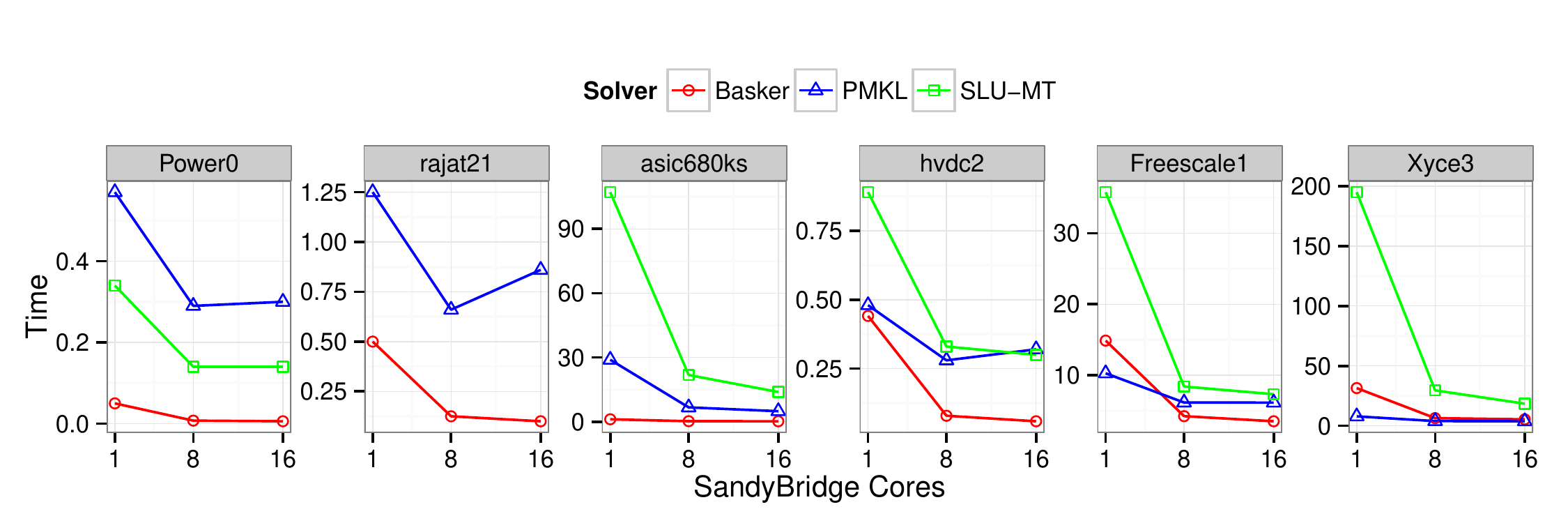}
  \caption{Comparison of \basker, PMKL, and SLU-MT raw time (seconds) on SandyBridge. SLU-MT only does better on Power0 and fails on rajat21. }
  \label{fig:superlumt}
\end{figure}

We first compare the general performance of the chosen sparse solver packages.
Only the numeric time is compared, since the symbolic factorization of both \basker and PMKL is limited by finding \nd ordering.
Figure~\ref{fig:superlumt} gives the raw time on Intel SandyBridge for a selection of six matrices.
These six matrices are selected due to their varying fill-in density, and ordered increasing from a density of $1.3$ to $9.2$. 
We first observe that PMKL is as good or better than SuperLU-MT. Similar results have been reported in the past~\cite{vecparshylu} in comparing against SuperLU-Dist for circuit problems.
Additionally, \basker performs better than other solvers in 5/6 matrices.
For this reason, we only perform additional comparison to PMKL.

\subsection{Scalability}
In this section, we focus on the scalability of the numeric factorization phase of \basker and PMKL on the two architectures.
We use the relative speedup to KLU as that is the state-of-the-art sequential
solver, i.e., $Speedup(matrix,solver,p) =
\frac{Time(matrix,KLU,1)}{Time(matrix,solver,p)}$, where $Time$ is the time of
the numeric factorization phase, $matrix$ is the input matrix, $solver$ is
either \basker or PMKL, and $p$ is the number of cores.

\begin{figure}[ht!]
  \centering
  \subfigure[]
  {
  	\includegraphics[width=0.49\textwidth]{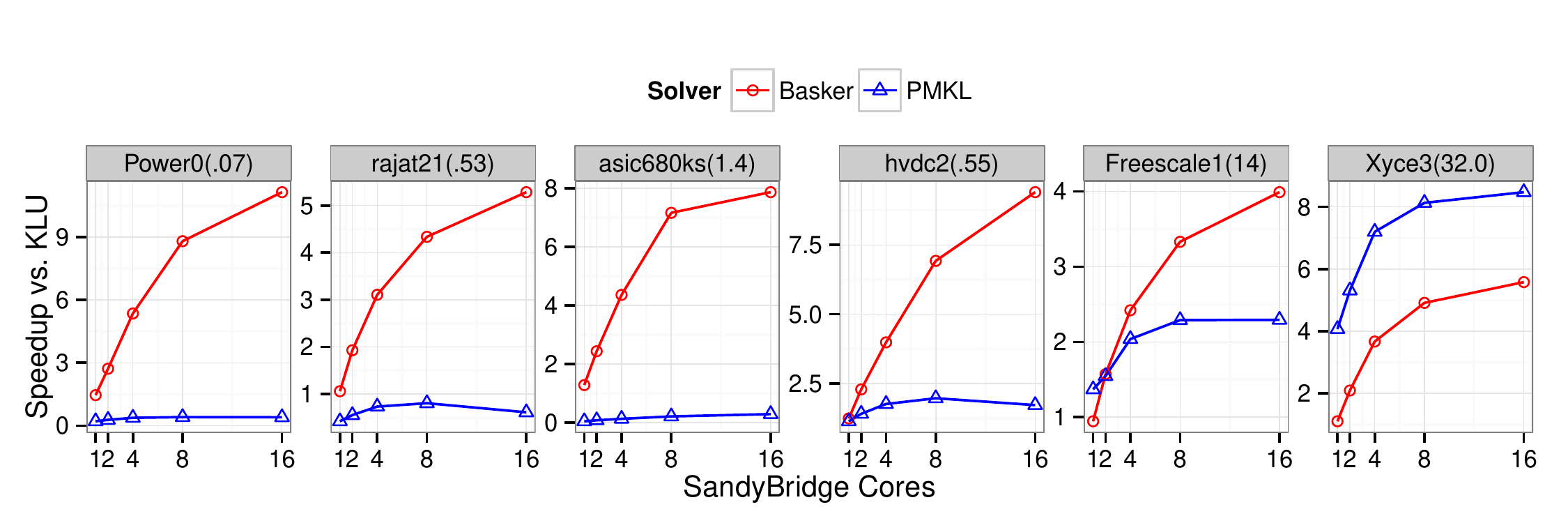}
  	\label{fig:speedupcpu}  
  }
  \vspace{-10pt}
  \subfigure[]
  {
    
    \includegraphics[width=0.49\textwidth]{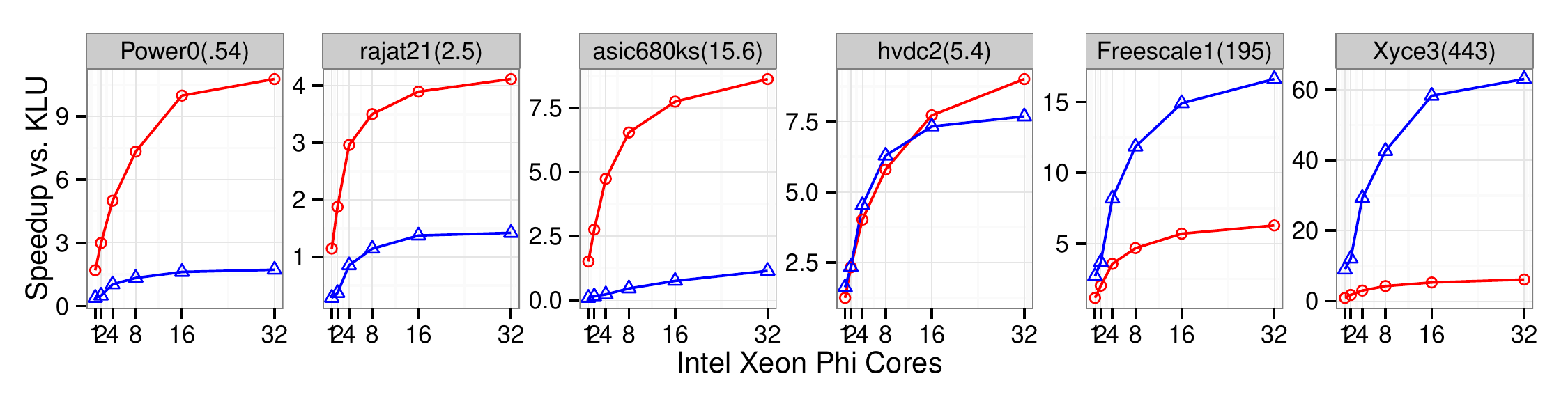}
    \label{fig:speedupphi}
   }
  \caption{Speedup of \basker and PMKL relative to KLU on Intel SandyBridge. KLU time is given in the title of each figure (a) and Xeon Phi (b) on six matrices that vary in fill-in density from low to high (left to right). Both Freescale1 and Xyce3 are considered to have high fill-in for \basker.  }
\end{figure}

Figure~\ref{fig:speedupcpu} shows the speedup achieved for these six matrices on SandyBridge platform.
We provide $Time(matrix,KLU,1)$ in the title of each figure. 
We observe that \basker can achieve up to $11.15\times$ speedup (hvdc2) and outperform PMKL in all but one case (Xyce3) that has a high fill-density of 9.2.
Moreover, we observe that PMKL has a speedup less than 1 in serial for four problems demonstrating the inefficiency of a supernodal algorithm to a \lgp for matrices with low fill-in density.
By adding more cores, PMKL is not able to recover from this inefficiency and reaches a max speedup of $2.34\times$ on the first four problems.
The reason for this is due to semi-dense columns that \basker is able to avoid factoring. 
PMKL does factor Xyce3 faster with its high fill-in density, but \basker scales
in a similar way.

The relative speedup of the same six matrices on the Intel Xeon Phi are shown in Figure~\ref{fig:speedupphi}.
Again, KLU time is provided in each figure's title.
On Intel Xeon Phi, \basker is able to out perform PMKL on four out of the six matrices. 
\basker achieves a $10.76\times$ maximum speedup (Power0) on these six matrices and PMKL achieves $63\times$ maximum speedup (Xyce3).
We observe that any overhead from using a \lgp on a matrix with high fill-in density is magnified by the Intel Phi.
This is exposed and seen in both Freescale1 and Xyce3.
One possible reason for this is that the submatrices in the lowest level of the hierarchical structure are too large to fit into a core's L2 cache ($512KB$).
\basker currently makes the submatrices as large as possible to allow for better pivoting. 
However, \basker still achieves speedups higher than PMKL on the four matrices with low fill-in density. 

\begin{figure*}[!th]
	\centering
	\subfigure[]
	{
		\includegraphics[width=.315\textwidth,height=.23\textwidth]{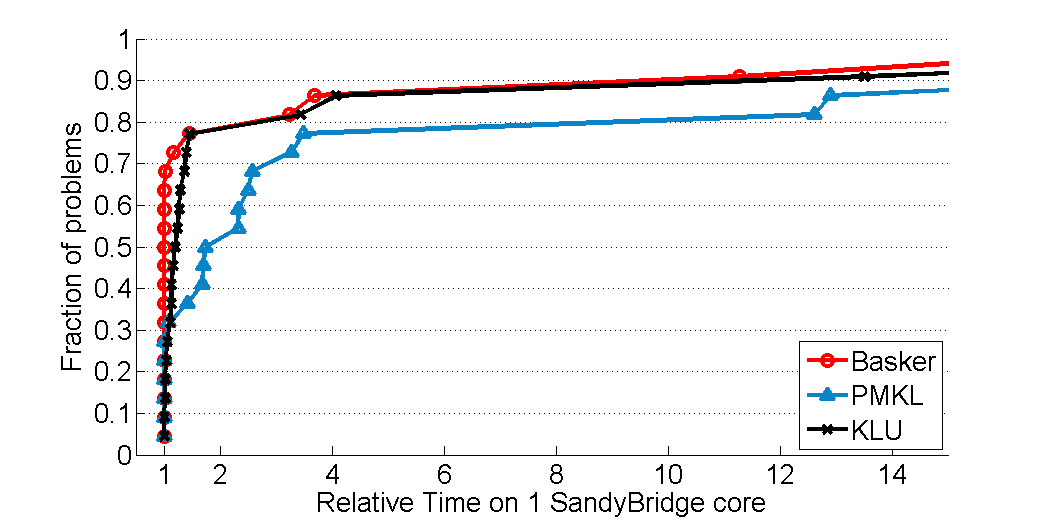}
		\label{fig:pp1cpu}
	}
	\vspace{-5pt}
	\subfigure[]
	{
		\includegraphics[width=.315\textwidth,height=.23\textwidth]{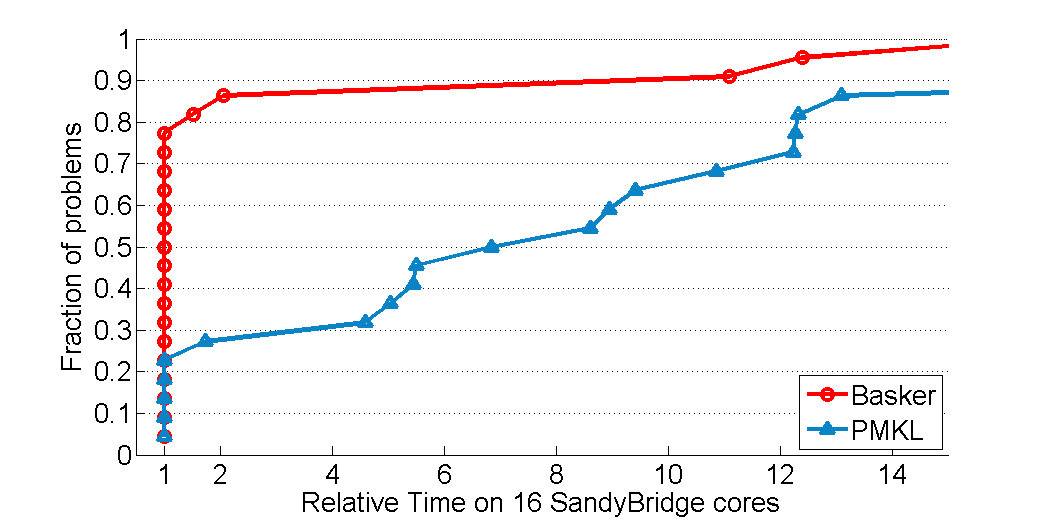}
		\label{fig:pp16cpu}
	}
	\vspace{-5pt}
	\subfigure[]
	{
		\includegraphics[width=.315\textwidth,height=.23\textwidth]{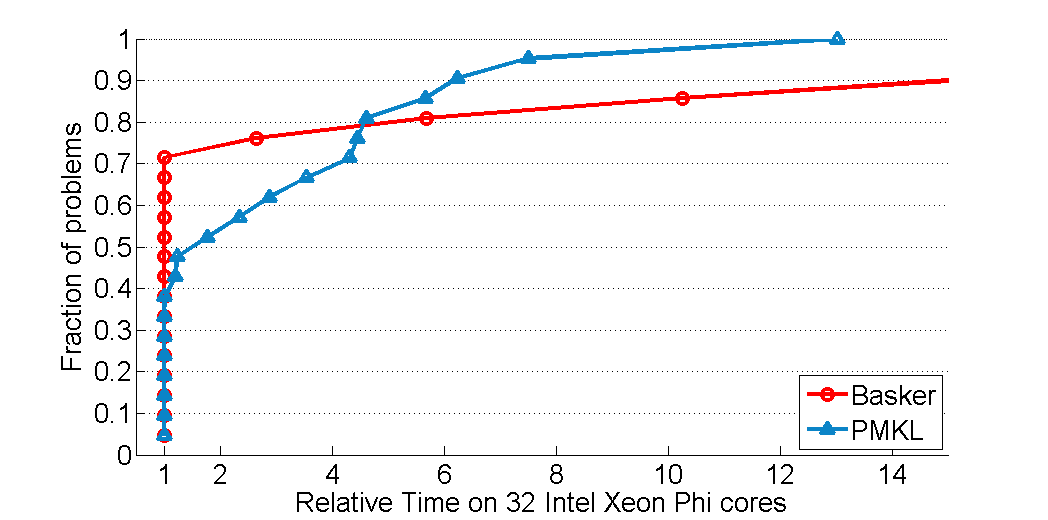}
		\label{fig:pp32phi}
	}	
	\caption{Performance profiles of \basker and PMKL on Intel SandyBridge and Xeon Phi.  A point (x,y) represents the fraction y of test problems within x$\times$ of the best solver. (a) 1 SandyBridge Core. \basker is the best solver for almost 70\% of the matrices and PMKL is the best solver for about $30\%$. (b) 16 SandyBridge Cores. \basker is the best solver for almost $80\%$ of the matrices, while PMKL is the best solver for slightly more than $20\%$. (c)32 Phi Cores. \basker is the best solver for over $70\%$ of the matrices, while PMKL is the best solver for $40\%$ of the matrices.}
\end{figure*}

As a next step, we compare the performance on the whole test suite.
On SandyBridge, the geometric mean of speedup for all the matrices with \basker is  $5.91\times$ and with PMKL is it $1.5\times$ using 16 cores.
On 16 cores, \basker is faster than PMKL on 17/22 matrices. The five matrices PMKL is faster on have a high fill-in density.
On the Xeon Phi, the geometric mean speedup with \basker is $7.4\times$ and with PMKL it is $5.78\times$ using 32 cores.
On 32 cores, \basker is faster than PMKL on 16/22 matrices. 
This includes the same matrices as on the SandyBridge except Freescale1.
The reason for such a high speedup for PMKL on Xeon Phi is again its higher performance on high fill-in density matrices.

 While the geometric mean gives some idea on relative performance, we use a
performance profile to gain an understanding of the overall performance over the
test suite.  The performance profile measures the relative time of a solver on
a given matrix to the best solver.  The values are plotted for all matrices in
a graph with an x-axis of time relative to best time and a y-axis as fraction
of matrices.  The result is a figure where a point(x,y) is plotted if a solver
takes no more than x times the runtime of of the fastest solver for y problems.

Figure~\ref{fig:pp1cpu} shows the performance profile of \basker, PMKL, and KLU \emph{in serial} on SandyBridge.
This shows a baseline of how well each method does in serial.
We observe that \basker is better on $\sim 77\%$ of the problems, while the supernodal method of PMKL is within $5\times$  of the the best solver for $77\%$ of 
the problems.
However, PMKL is the better solver for $\sim 34\%$ of the problems.
Despite having very similar algorithms, \basker is able to slightly beat KLU.
This slight difference is because of the difference in orderings and the use of Kokkos memory padding.

The performance profile of the parallel solvers on SandyBridge (16 cores) is
shown in Figure~\ref{fig:pp16cpu}.  Serial KLU is not included in this figure.
\basker is the best solver for $\sim 75\%$ of the matrices, and PMKL is within
$\sim5\times$  of \basker on $\sim 50\%$ of the matrices.  PMKL is the best
solver for $\sim 30\%$ of the matrices, which correspond to matrices with high
fill-in density.  This demonstrates \basker scales well on SandyBridge for low
fill-in density matrices.
On Intel Xeon Phi with 32 cores, the performance profile is slightly different (Figure~\ref{fig:pp32phi}).
\basker now is the best solver for $70\%$ matrices, and PMKL is within $6\times$ of \basker for $70\%$ of matrices.   
PMKL is the best (or very close to the best) for $\sim40\%$ of the matrices.
One can observe \basker now does poorly on high fill-in density matrices.
A reason for that is the missing large shared $L3$ to share data needed during the \basker's reductions.

\subsection{Comparison on Ideal Matrices}
Next, we analyze how well \basker scales on low fill-in density matrices, compared to how well the supernodal solver PMKL scales on 2/3D mesh problems.
This comparison allows us to better understand if \basker achieves speedup for its ideal input similar to PMKL on its ideal input. The other reason is to see
how well we can parallelize \lgp for its ideal problems.
We use a second test suite of matrices for PMKL that come from 2/3D mesh problems in Table II. 
 Performance of PMKL on these matrices will be compared to the performance of \basker on the six matrices of our primary test suite with the lowest fill-in density.

\begin{table}[htbp]
  \centering
	{\verysmallfont
	\tabcolsep=0.08cm
  \caption{2/3D mesh problems to test PMKL's best performance.}
    \begin{tabular}{c|c|c|c|c}
		\hline
    Matrix         & $n$    & $|A|$ & $|L$+$U|$ & Description  \\
		\hline
    pwtk           & 2.2E5  & 1.2E7 & 9.7E7 & Wind tunnel stiffness matrix \\
    ecology        & 1.0E6  & 5.0E6 & 7.1E7 & 5 pt stencil model movement \\
    apache2        & 7.2E5  & 4.8E6 & 2.8E8 & Finite difference 3D \\
    bmwcra1        & 1.5E5  & 1.1E7 & 1.4E8 & Stiffness matrix \\
    parabolic\_fem & 5.3E5  & 3.7E6 & 5.2E7 & Parabolic finite element \\
    helm2d03       & 3.9E5  & 2.7E6 & 3.7E7 & Helmholtz on square \\
		\hline
    \end{tabular}%
		}
  \label{tab:pideal}%
\end{table}%

\begin{figure}[h]
	\centering
	\subfigure[]
	{
		\includegraphics[width=0.22\textwidth]{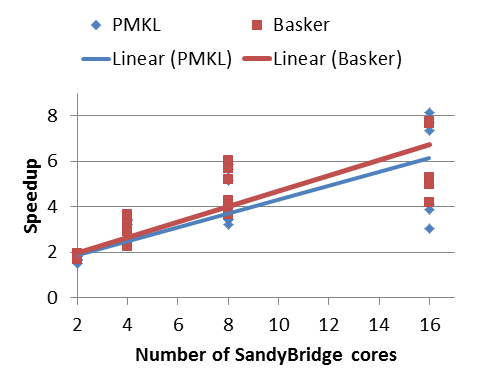}
		\label{fig:ideal_cpu}
	}
	\subfigure[]
	{
		\includegraphics[width=0.22\textwidth]{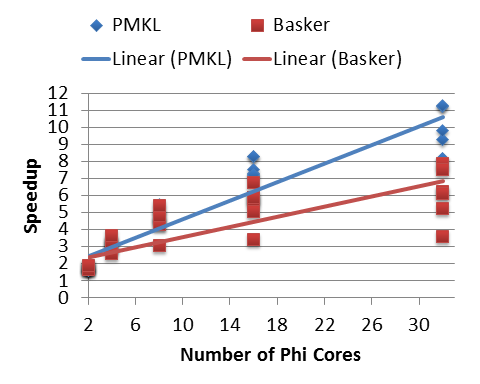}
		\label{fig:ideal_phi}
	}
	\caption{\basker and PMKL with on 6 ideal input matrices. (a) SandyBridge, \basker is able to achieve a similar speedup curve as PMKL on 2/3D mesh problems. (b) Intel Xeon Phi, \basker has a similar plot up to 16 cores as PMKL.  Fine-grain access causes imbalance at 32 cores. }
\end{figure}

Figure~\ref{fig:ideal_cpu} provides a scatter plot of the speedup for each solver relative to itself over its ideal six matrices.
A linear trend line is shown for each set of solver speedups.
Both solvers achieve  similar speedup trend on SandyBridge for
their ideal inputs.
This demonstrates that on systems with a large cache hierarchy \basker is able to achieve so called state-of-the-art performance on low fill-in density matrices.
In Figure~\ref{fig:ideal_phi}, a similar plot is given for our Xeon Phi platform.
This time \basker has a slightly lower trend line starting at 16 cores.
We suspect this is due to both the size of the submatrices not fitting into cache and time for the reduction.
We plan to address both these issues in future versions of \basker.

\subsection{Xyce}

Next, we consider the use of \basker on a sequence of matrices generated during the transient analysis of a circuit.
Xyce is a transistor-level simulator that performs a SPICE-style simulation of circuits, where devices and their interconnectivity are transformed
via modified nodal analysis into a set of nonlinear differential algebraic equations (DAEs).  During transient analysis, these nonlinear DAEs
are solved implicitly through numerical integration methods.  Any numerical integration method requires the solution to a sequence of nonlinear
equations, which in-turn generates a sequence of linear systems.  A transient analysis can generate millions of coefficient matrices with the 
same structure and significantly different values.  Each factorization may require a different permutation due to pivoting for this reason.
For very large circuits, this results in the numeric factorization being the limiting factor of the simulation overall time and scalability.
Furthermore, a solver package must reuse the symbolic factorization for all matrices in the sequence as repeating symbolic 
factorization would dramatically affect performance. 

For this experiment, we chose a sequence from the circuit that generated Xyce1.  This circuit is of particular interest because it has been 
used in prior studies~\cite {ICCAD09_precond} to illustrate the ineffectiveness of preconditioned iterative methods and direct solvers other than KLU.  
Attempts to use the PMKL solver had either been met with solver failure or simulation failure until recently.    
Therefore, we wish to see how well \basker performs on a sequence of these matrices (1000 matrices) which represent $10\%$ of the desired transient length.
 
Over the sequence of 1000 matrices, \basker took 175.21 seconds, KLU took 914.77 seconds, and PMKL took 951.34 seconds.  
This is a speedup of $5.43\times$ when using \basker instead of PMKL and $5.22\times$ when using \basker instead of KLU.   
The scalable simulation of this circuit was previously limited by the serial bottleneck produced by using KLU as the direct solver, 
which is justified due to its performance compared to PMKL.
\basker provides significant speedup compared to either KLU or PMKL, and will finally provide a scalable direct solver to Xyce for performing the
transient analysis of this circuit.

\section{Conclusions and Future Work}
\label{sec:conclusion}

We introduced a new multithreaded sparse $LU$ factorization, \basker, that uses hierarchical parallelism and data layouts.
\basker provides a nice alternative to traditional solvers that use one-dimensional layout with BLAS.
In particular, it is useful for coefficient matrices with hierarchical structure such as circuit problems.
We also introduced the first parallel implementation of \lgp.
Performance results show that \basker scales well for matrices with low fill-in density resulting in a speedup of $5.91\times$ (geometric mean) over the test suite on 16 SandyBridge cores and $7.5\times$ over the test suite on 32 Intel Xeon Phi cores relative to KLU.
Particularly, \basker can have speedups on these matrices similar to PMKL on 2/3D mesh problems and reduce the time for a sequence of circuit problems from Xyce by $5\times$.
\basker shows that in order to speedup sparse factorization on many-core node, solvers must leverage all available parallelism and may do so by using a hierarchical structure.

We plan to continue support of \basker in the ShyLU package of Trilinos for Xyce.
Future scheduled improvements include adding supernodes to the hierarchy structure to improve performance on high fill-in matrices, and using asynchronous tasking to reduce synchronization costs.

\section*{Acknowledgment}
We would like to thank Erik Boman, Andrew Bradley, Kyungjoo Kim, H.C. Edwards, Christian Trott, and Simon Hammond for insights and discussions. 
Sandia is a multiprogram laboratory operated by Sandia Corporation, a
Lockheed Martin Company, for the U.S. Department of Energy under
contract DE-AC04-94-AL85000.

\bibliographystyle{IEEEtran}
\bibliography{Ref}

\end{document}